\documentclass[twocolumn]{aastex631}

\usepackage[T1]{fontenc}
\usepackage{amsmath}
\usepackage{enumitem}
\usepackage{natbib}
\usepackage{color,soul}
\usepackage{afterpage}

\newcommand{\Lya}{Ly-$\alpha$}

\begin{document}

\title{Resolving Clumpy vs. Extended \Lya\ In Strongly Lensed, High-Redshift \Lya\ Emitters}

\shorttitle{Resolving Clumpy vs. Smooth \Lya\ In Strongly Lensed, High-Redshift \Lya\ Emitters}
\shortauthors{Navarre et al.}

\author[0000-0001-7548-0473]{Alexander Navarre}
\affiliation{Department of Physics, University of Cincinnati, Cincinnati, OH 45221, USA}

\author[0000-0002-3475-7648]{Gourav Khullar}
\affiliation{Department of Physics and Astronomy and PITT PACC, University of Pittsburgh, Pittsburgh, PA 15260, USA}

\author[0000-0003-1074-4807]{Matthew B. Bayliss}
\affiliation{Department of Physics, University of Cincinnati, Cincinnati, OH 45221, USA}

\author[0000-0003-2200-5606]{H\aa kon Dahle}
\affiliation{Institute of Theoretical Astrophysics, University of Oslo, Oslo, Norway}

\author[0000-0001-5097-6755]{Michael Florian}
\affiliation{Steward Observatory, University of Arizona, Tuscon, AZ 85719, USA}

\author[0000-0003-1370-5010]{Michael Gladders}
\affiliation{Department of Astronomy and Astrophysics, University of Chicago, Chicago, IL 60637, USA}

\author[0000-0001-6505-0293]{Keunho J. Kim}
\affiliation{Department of Physics, University of Cincinnati, Cincinnati, OH 45221, USA}

\author[0000-0002-2862-307X]{M. Riley Owens}
\affiliation{Department of Physics, University of Cincinnati, Cincinnati, OH 45221, USA}

\author[0000-0002-7627-6551]{Jane Rigby}
\affiliation{NASA Goddard Space Flight Center, Greenbelt, MD 20771, USA}

\author[0000-0002-0975-623X]{Joshua Roberson}
\affiliation{Department of Physics, University of Cincinnati, Cincinnati, OH 45221, USA}

\author[0000-0002-7559-0864]{Keren Sharon}
\affiliation{Department of Astronomy, University of Michigan, Ann Arbor, MI 48109, USA}

\author{Takatoshi Shibuya}
\affiliation{Kitami Institute of Technology, Kitami, Hokkaido, Japan}

\author[0000-0001-5424-3698]{Ryan Walker}
\affiliation{Department of Astronomy, University of Michigan, Ann Arbor, MI 48109, USA}

\begin{abstract}
We present six strongly gravitationally lensed \Lya\ Emitters (LAEs) at $z\sim4-5$ with HST narrowband imaging isolating \Lya. Through complex radiative transfer \Lya\ encodes information about the spatial distribution and kinematics of the neutral hydrogen upon which it scatters. We investigate the galaxy properties and \Lya\ morphologies of our sample. Many previous studies of high-redshift LAEs have been limited in \Lya\ spatial resolution. In this work we take advantage of high-resolution \Lya\ imaging boosted by lensing magnification, allowing us to probe sub-galactic scales that are otherwise inaccessible at these redshifts. We use broadband imaging from HST (rest-frame UV) and Spitzer (rest-frame optical) in SED fitting; providing estimates on the stellar masses ($\sim 10^8 - 10^9 M_{\odot}$), stellar population ages ($t_{50} <40$ Myr), and amounts of dust ($A_V \sim 0.1 - 0.6$, statistically consistent with zero). We employ non-parametric star-formation histories to probe the young stellar-populations which create \Lya. We also examine the offsets between the \Lya\ and stellar continuum, finding small upper limits of offsets ($<$ 0\farcs1) consistent with studies of low-redshift LAEs; indicating our galaxies are not interacting or merging. Finally, we find a bimodality in our sample's \Lya\ morphologies: clumpy and extended. We find a suggestive trend: our LAEs with clumpy \Lya\ are generally younger than the LAEs with extended \Lya, suggesting a possible correlation with age. 
\end{abstract}


\section{Introduction} \label{sec:intro}
\subsection{Lyman-\texorpdfstring{$\alpha$}{a} and LAEs} \label{subsec:Lyman-alpha}
Lyman Alpha (herein Ly-$\alpha$) is emitted from the atomic transition $n = 2$ to $n = 1$ of hydrogen. It is hydrogen's brightest recombination line, and abundantly appears in many young, star-forming galaxies. Due to its complex interactions with neutral hydrogen gas, \Lya\ is an important tool for studying the properties of young stellar populations. It traces recent star formation and carries information about the morphology and kinematics of neutral hydrogen gas.

Young, massive stars in regions of neutral hydrogen (H\,I regions) produce \Lya\ in abundance. To first order, these stars ionize their surroundings, creating regions of ionized hydrogen (H\,II regions). At the boundary of the H\,I and H\,II regions the ionization rate and recombination rate balance out. As recombination happens the electron cascades down to the ground state, with a likelihood of $\approx$ 68\% to create a \Lya\ photon (this assumes Case B recombination and a temperature of \linebreak $10^{4}$ K; see \citealt{Dijkstra2017}). This can be thought of as an effective conversion factor which locally converts ionizing photons into \Lya\ photons. Although, it is important to note that other mechanisms, such as collisional excitation, also create \Lya\ photons within these environments. Significant \Lya\ emission traces the formation of these stars over cosmological timescales, implying that the host galaxy is young. Similarly, it implies that there is not an abundance of dust in the host galaxy, which would otherwise attenuate the \Lya\ flux \citep{Scarlata2009,Henry2015,Saldana2023}.

A class of galaxies exists which emit significant amounts of \Lya\ \citep{Cowie1998}. These galaxies are called \Lya\ Emitters (LAEs), and are typically classified as having some minimum EW. This minimum EW is often related to the detection thresholds of the narrowband imaging surveys used to identify samples of LAEs. There is no obvious consensus in the literature on a single Lya EW threshold, though EW $\gtrapprox 20$ \AA\ is perhaps the most common. Our sample all have Lya EW $>15$ \AA, which is very similar (albeit slightly below) the most common literature selection criteria. LAEs are thought to be young and actively star-forming with low to moderate dust content \citep{Gronwall2007,Finkelstein2008,Ouchi2008}. 

\subsection{Lyman-\texorpdfstring{$\alpha$}{a} Radiative Transfer} \label{subsec:Radiative Transfer}
There is a complex set of radiative transfer processes that \Lya\ photons can undergo with neutral hydrogen  \citep{Dijkstra2017}. The simplest conceptual path for a \Lya\ photon to take is direct escape from the galaxy after its initial creation. However, we see in many star-forming galaxies that the \Lya\ emission is spatially extended and/or offset from the associated stellar continuum \citep{Ostlin2009,Hayes2013}. This happens because the photons interacted with neutral hydrogen atoms in-between their initial creation and escape. Neutral hydrogen interacting with a \Lya\ photon will absorb and re-emit the photon in a random direction, effectively scattering it. This scattering process can happen many times before the \Lya\ photon is able to escape the galaxy, allowing the \Lya\ photon to travel far from its original creation site. This leads to \Lya\ emission that is spatially broader than the stellar continuum.

\subsection{High-Redshift LAEs} \label{subsec:High z}

LAEs are typical targets for high-redshift galaxy studies because of their excess brightness in narrowband imaging. Many LAEs have been found at low \citep{Deharveng2008,Cowie2011,Ostlin2014} and high redshifts \citep{Cowie1998,Rhoads2000,Yamada2005,Ouchi2008,Marques-Chaves2017,Mukae2020,Kikuta2023}, suggesting that they are galaxies in a specific stage of evolution; thought to be the progenitors of Milky Way type galaxies in the local universe \citep{Ono2010,Dressler2011,Guaita2011}.  However, the astrophysical mechanisms that differentiate this stage are not well understood at high redshifts. For example, narrowband \Lya\ imaging of LAEs shows heterogeneous morphologies. They range from clumpy \citep[scale radii $\sim$ 0.5 kpc;][]{Bond2010,Finkelstein2011} and comparable to stellar continuum sizes, to moderately extended \citep[scale radius $\sim$ 2 kpc;][]{Jung2023},  to significantly extended \citep[scale radii $\sim$ 80 kpc;][]{Steidel2011}.

Recent studies with JWST and MUSE have shown the importance of faint LAEs to cosmic reionization. \citet{Thai2023} finds an abundance of faint LAEs at high redshift, and \cite{Atek2023} finds that faint high-redshift LAEs have very high ionizing photon production efficiencies. This suggests that LAEs contribute more than previously thought to cosmic reionization. \citet{Witstok2023} finds that their sample of spectroscopically-confirmed LAEs at $z > 7$ are alone not sufficient to produce the ionized bubble sizes inferred from their spectra. They suggest that ultra-faint ($M_{UV} \gtrsim -18$) LAEs likely play an important role in carving out these bubbles. \citet{HA2023} finds that undetected (presumably faint) LAEs could dominate excess surface brightness seen at large scales in \Lya\ halos. Magnification from gravitational lensing will be an essential tool in future studies of very faint LAEs during the epoch of reionization.

The goal of this paper is to characterize a sample of high-redshift, gravitationally lensed LAEs imaged by HST and Spitzer. We measure their stellar mass, age, and dust content ($A_{V}$), gaining context on the environment in which the \Lya\ emitting regions exist. We also measure the \Lya\ flux and luminosity in both the image plane (observed, magnified) and source plane (intrinsic, demagnified). These measurements, in conjunction with previously measured \Lya\ equivalent widths, show the strength of the \Lya\ emission. 

This paper is organized as follows: in section 2 we introduce our sample of LAEs and summarize the data used. In section 3 we describe our methodologies and measurements. In section 4 we analyze our measurements and discuss our findings. Throughout this paper we assume a $\Lambda$CDM cosmology with ($H_0$, $\Omega_m$, $\Omega_\Lambda$) = (67.7 kms$^{-1}$ Mpc$^{-1}$, 0.31, 0.69).

\begin{figure*}[!htb]
    \centering
    \includegraphics[width=\textwidth]{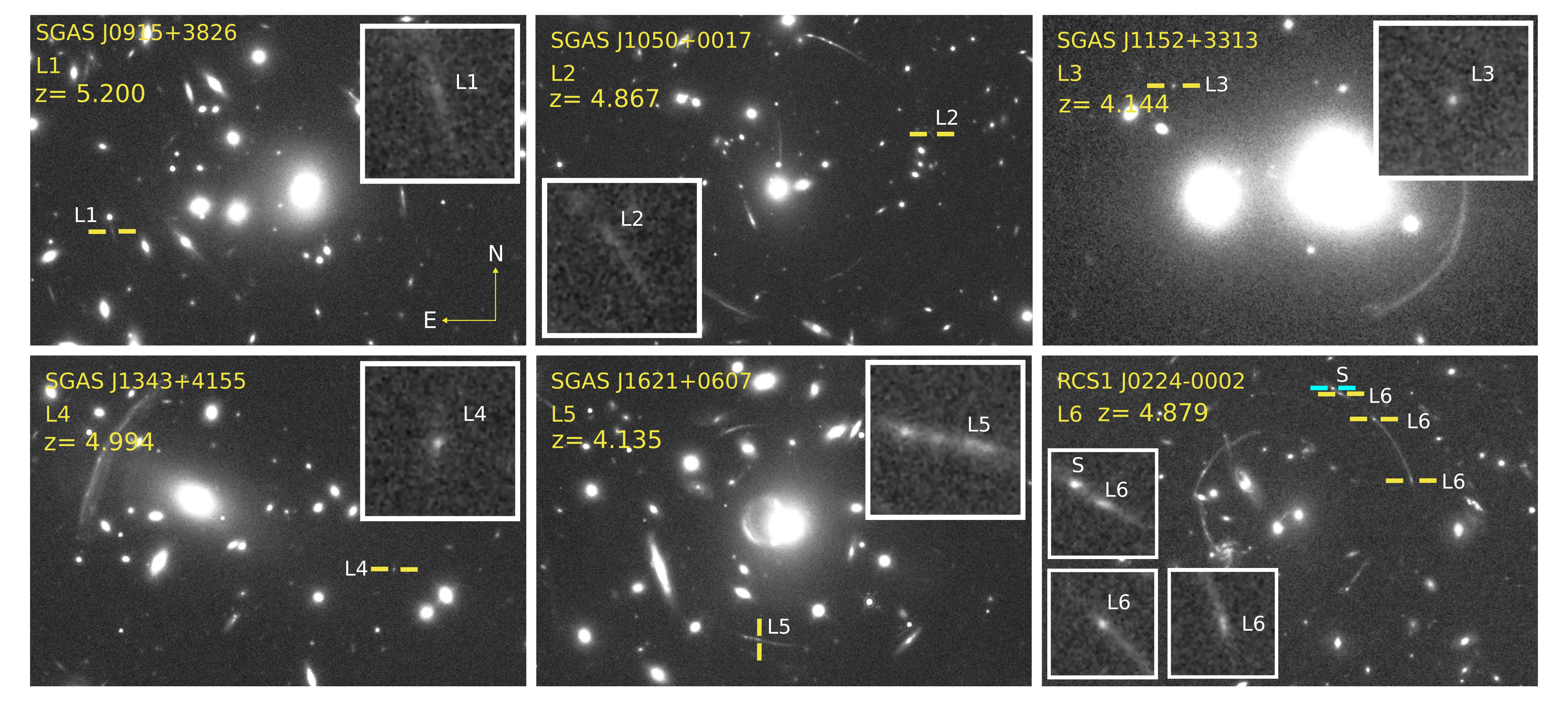}
    \figcaption{The LAE sample and lensing galaxy clusters. In the cluster view we show F160W (F814W for L6) with the LAEs located by sets of yellow bars. In the postage stamps, close up views of the LAEs are shown in the bluest available filter redward of \Lya\ that does not contain \Lya. These filters are reported in \autoref{tab:offsets} and capture light from the stellar continuum. L6 contains three bright, separated images. A bright galaxy is seen projected onto L6, which we label S and indicate with a set of cyan bars. This galaxy is not gravitationally lensed nor part of L6, but is instead a foreground $z=3.66$ galaxy (identified by \citet{Swinbank2007}) that serendipitously appears to be part of the arc. \label{fig:view}}
    
\end{figure*}

\begin{deluxetable*}{c c c c c c c c}[!htb]
    \startdata \\
    Identifier & Lensing Cluster & Cluster & LAE & Cluster & Cluster & LAE & LAE\\
     & & Redshift & Redshift & RA & Dec & RA & Dec\\
    \hline\hline
    L1a & SGAS J091541+382655 & 0.396 & 5.200 & 09:15:39 & +38:26:59   & 09:15:41.0 & +38:26:53.2 \\
    L1b & -- & -- & -- & -- & -- & 09:15:41.0 & +38:26:52.3 \\
    L2a & SGAS J105038+001715 & 0.593 & 4.867 & 10:50:40 & +00:17:07   & 10:50:38.3 & +00:17:14.9\\
    L2b & -- & -- & -- & -- & -- & 10:50:38.4 & +00:17:15.2\\
    L3 & SGAS J115201+331347 & 0.517 & 4.144 & 11:52:00  & +09:30:15  & 11:52:01.0 & +33:13:48.2 \\
    L4 & SGAS J134331+415455 & 0.418 & 4.994 & 13:43:33 & +41:55:04   & 13:43:30.7 & +41:54:55.1 \\
    L5 & SGAS J162132+060705 & 0.343 & 4.135 & 16:21:32 & +06:07:20   & 16:21:32.6 & +06:07:05.6 \\
    L6$'$ & RCS1 J022434-000220 & 0.773 & 4.879 & 02:24:34 & -00:02:31   & 02:24:33.8 & -00:02:17.8 \\
    L6$''$ & -- & -- & -- & -- & -- & 02:24:33.6 & -00:02:20.5 \\
    L6$'''$ & -- & -- & -- & -- & -- & 02:24:33.3 & -00:02:26.9 \\
    \enddata
    \caption{Positions and redshifts of lensed LAEs and lensing clusters. \citep{Gladders2002,Sharon2020}. Lowercase letters denote bright regions inside a single image, while apostrophes denote different images of the same bright region. See \autoref{subsec:Dist} and \autoref{fig:roguesgallery} for justification for analyzing our LAEs in this manner. Additionally, we investigate only the brightest and/or most isolated images in each case. We refer the reader to \citet{Gladders2002,Sharon2020} for information on the other lensed images.}
    \label{tab:position}
\end{deluxetable*}

\begin{figure*}[ht]
    \centering
    \includegraphics[scale=0.5]{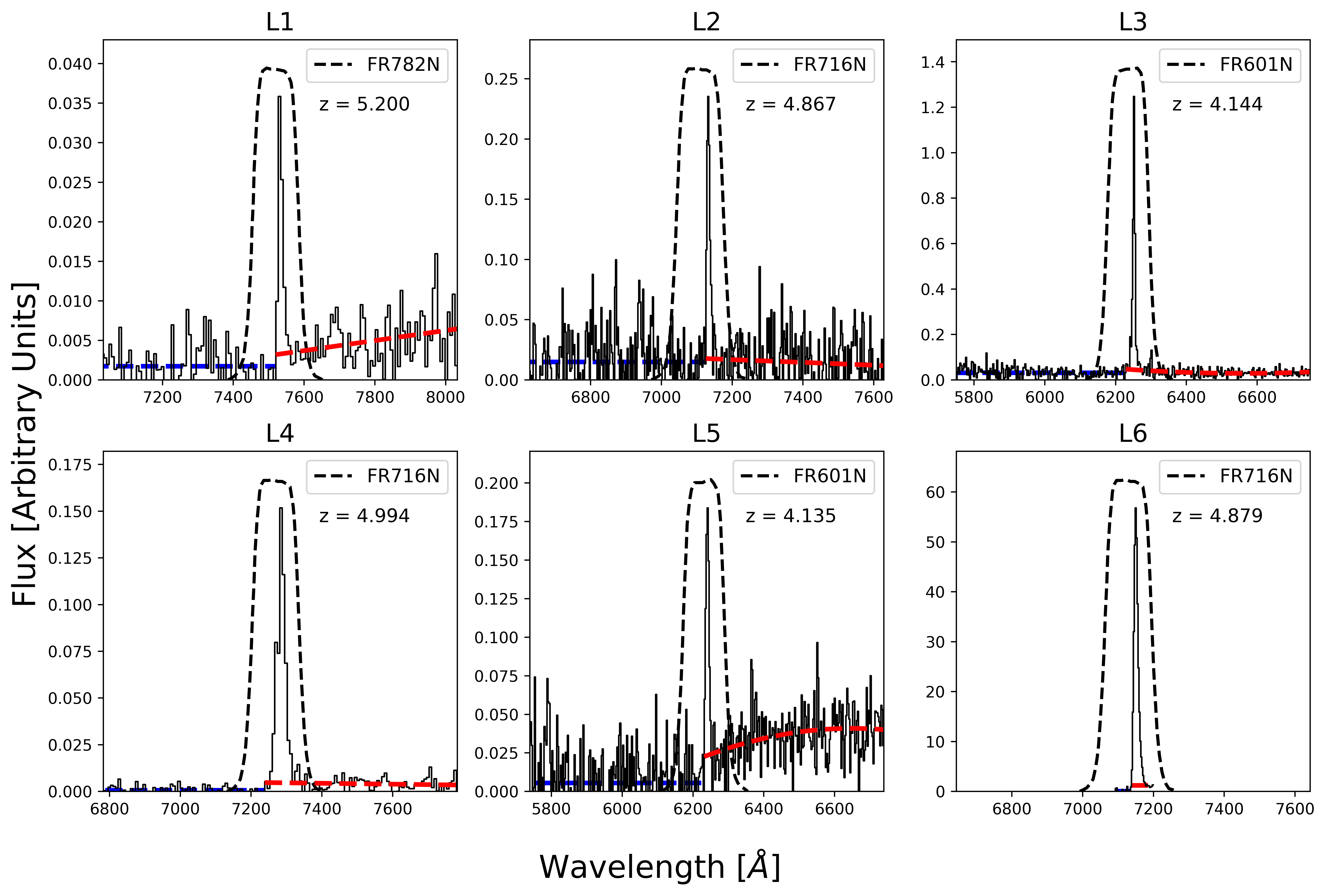}
    \figcaption{GMOS (L1-L5) and VIMOS IFU (L6) spectra of the LAE sample centered at \Lya. The dashed black curve represents the throughput of the ramp filter used to capture \Lya. The dashed red and blue curves represent fits to the continuum levels redward and blueward of \Lya. When calculating \Lya\ fluxes and luminosities we include a multiplicitive factor accounting for the fraction of light in the narrowband image coming directly from the \Lya\ line. The ramp filter properties can be found in Table \ref{tab:filters}.  \label{fig:spec} }
   
\end{figure*}

\section{Data} \label{sec:Data}

\subsection{Lensed LAE Sample} \label{subsec:Data Sample}

We analyze a sample of six high-redshift ($z>4$) LAEs. These six lensed LAEs are in the redshift range $4.1 < z < 5.2$. All six have published, well-constrained strong lensing models \citep{rzepecki2007,Swinbank2007,Bayliss2010,Bayliss2014,Smit2017,Sharon2020}, allowing for spatial analysis at resolutions otherwise inaccessible at these redshifts. Previous studies such as \citet{Marques-Chaves2017,Berg2018,Erb2019} have shown the utility of lensing magnification in characterizing the properties of \Lya\ emitting galaxies both in the image plane and source plane.

All six lensed LAEs have imaging in broadband filters from HST (rest-frame UV) 
and Spitzer (rest-frame optical), allowing for broad constraints on the stellar 
population properties in each galaxy. In this paper we present new narrow-band imaging using tuneable  
ramp filters on HST that isolate \Lya. These filters allow us to directly measure the morphology and spatial extent of the \Lya\ emission. The individual sources and imaging 
data sets are described in more detail below. For ease of reference, our sample of six LAEs will 
be referred to as L1--L6, as shown in Figure \ref{fig:view} and Tables \ref{tab:position} \& \ref{tab:lyaprop}.

\begin{deluxetable}{c c c c c c}[!htb]
    \startdata
    \\
    ID & Ramp & \Lya\ & Central & Exposure & Detection\\
     & Filter & [\AA] & $\lambda$ [\AA] & Time [s] & Limit [$\times 10^{-16}$]\\
    \hline\hline
    L1 & FR782N & 7533.0 & 7510 & 7840 & 1.21\\
    L2 & FR716N & 7158.4 & 7134 & 5060 & 0.25\\
    L3 & FR601N & 6249.9 & 6235 & 7800 & 0.33\\
    L4 & FR716N & 7282.7 & 7270 & 5260 & 0.44\\
    L5 & FR601N & 6239.0 & 6230 & 7700 & 0.49\\
    L6 & FR716N & 7142.9 & 7149 & 2400 & 0.53\\
    \enddata
    \caption{Description of the HST narrowband ramp filter observations. These filters were chosen for their ability to adjust their wavelength coverage. The central wavelengths of the ramp filters were offset to cover the Lyman-Alpha lines seen in the associated Gemini and VLT spectra (see \autoref{fig:spec}). Detection limit refers to the 1-$\sigma$ uncertainty in the sky background level within an aperture covering a solid angle of 1 square arcsecond. The detection limit is reported in ergs/s/cm$^2$/arcsec$^2$.}
    \label{tab:filters}
\end{deluxetable}

\subsection{Sample Discovery and Spectroscopic Observations} \label{subsec:Spec}

The sample of strongly lensed \Lya\ emitting galaxies are drawn from the literature. 
L6 was first identified in the Red-Sequence Cluster Survey \citep[RCS;][]{Gladders2003} as a highly extended ``giant arc'' around a massive galaxy cluster, RCS1~J0224-0002 (see \autoref{fig:view}). It was found to have a redshift of $z=4.8786$ based on strong \Lya\ emission observed with the FORS2 spectrograph on the VLT \citep{Gladders2002}. L6 has been the subject of several detailed studies using ground-based optical and NIR spectroscopy \citep{Swinbank2007,Smit2017,Witstok2021}. The spectra of \Lya\ seen in \autoref{fig:spec} is from the VIMOS IFU observations from \citet{Swinbank2007}, where the IFU covered the entire arc. 

\begin{deluxetable*}{c c|c c c c c c}[!ht]
    \startdata
    \\
    Telescope & Filter & L1 & L2 & L3 & L4 & L5 & L6 \\
    \hline\hline
    HST & F160W & 23.60 $\pm$  0.02 & 24.75 $\pm$  0.05 & 24.60 $\pm$  0.03 & 24.16 $\pm$  0.02 & 21.92 $\pm$  <0.01 & 22.43 $\pm$ 0.03 \\
    -- & F125W & 23.43 $\pm$  0.02 & -- & -- & -- & -- & 22.56 $\pm$ 0.02\\
    -- & F110W & -- & 24.45 $\pm$  0.02 & 24.58 $\pm$  0.06 & -- & 21.97 $\pm$  <0.01 & -- \\
    -- & F105W & -- & -- & -- & 24.32 $\pm$  0.03 & -- & -- \\
    \hline
    HST & F850LP & -- & -- & -- & -- & -- & 22.62 $\pm$  0.08 \\
    -- & F814W & 23.38 $\pm$ 0.02 & -- & -- & 24.16 $\pm$  0.04 & -- & 21.94 $\pm$  0.01 \\
    -- & F775W & -- & -- & -- & -- & 22.10 $\pm$  0.02 & -- \\
    -- & F606W & -- & 26.01 $\pm$  0.15 & 24.16 $\pm$  0.03 & 27.21 $\pm$ 0.19 & -- & 23.49 $\pm$  0.02 \\
    -- & F475W & -- & -- & 25.58 $\pm$ 0.10 & -- & -- & -- \\
    -- & F390W & $<\emph{7.45e-12}$ & $<\emph{1.64e-12}$ & -- & $<\emph{8.08e-14}$ & $<\emph{1.28e-11}$ & -- \\
    \hline
    Spitzer & Ch1 & 24.32 $\pm$ 0.97 & 23.76 $\pm$ 0.19 & 23.75 $\pm$ 0.62 & 23.51 $\pm$  0.39 & 20.77 $\pm$  0.02 & 21.31 $\pm$  0.04 \\
    -- & Ch2 & -- & 23.85 $\pm$  0.55 & $<\emph{1.80e-10}$ & -- & 20.89 $\pm$ 0.04 & 22.09 $\pm$  0.13  \\
    \hline
    Ground-Based & i-band & \textbf{23.34 $\pm$  0.09} & -- & -- & \textbf{23.78 $\pm$  0.18} & -- & -- \\
    --& z-band & \textbf{23.39 $\pm$ 0.13} & -- & -- & \textbf{24.24 $\pm$  0.17} & -- & -- \\
    \enddata
    \tablecaption{AB Magnitudes}
    \caption{Table of Broadband Photometry. We report the AB magnitudes and errors used in our analysis. The italicized entries are considered non-detections, and upper limits are reported in the SDSS unit of maggies to avoid very high AB magnitudes. Maggies are a dimensionless unit defined by: $\frac{f_{\nu}}{3631 \text{Jy}}$. Bold entries are measurements taken from \citet{Bayliss2010}. All other entries were calculated in this work.}
    \label{tab:phot}
\end{deluxetable*}

L1-5 are all located in strong lensing cluster fields found in the Sloan Digital Sky Survey \citep[SDSS;][]{York2000} (see \autoref{fig:view}). The sources were identified as g- or r-band dropout galaxies in $gri$ imaging from the GMOS-N instrument on the Gemini-North telescope. Redshifts for all these sources were measured from strong \Lya\ emission in follow-up spectroscopy with GMOS-N \citep{Bayliss2010,Bayliss2011,Bayliss2014}. The slits in these observations covered the integrated emission from individual images of each LAE. The redshifts for all six LAEs appear in Table~\ref{tab:position}

\begin{deluxetable*}{c c|c c c|c|c c c}[!ht]
    \startdata \\
     & & \multicolumn{3}{c|}{\textbf{Observed}}& & \multicolumn{3}{c}{\textbf{Intrinsic}}\\
    \hline
    ID & \Lya\ & UV & \Lya\ Flux & \Lya\ Luminosity & Magnification & UV & \Lya\ Flux & \Lya\ Luminosity\\
     & EW [\AA] & M$_{1500}$ & [$\times 10^{-18}$] & [$\times$ $10^{41}$] & Factor $\mu$ & M$_{1500}$ & [$\times 10^{-18}$] & [$\times$ $10^{41}$] \\
    \hline\hline
    L1 & 25 & -25.1 & 17.9 $\pm$ 7.6 & 53.3 $\pm$ 22.7 & 50 $\pm$ 17 &  -20.9 $\pm$ 0.4 & 0.4 $\pm$ 0.2 & 1.1 $\pm$ 0.6 \\
    L2 & 61 & -24.0 & 5.5 $\pm$ 0.9 & 13.9 $\pm$ 2.3 & 12.2 $\pm$ 0.7 &  -21.3 $\pm$ 0.06 & 0.4 $\pm$ 0.1 & 1.1 $\pm$ 0.2 \\
    L3 & 65 & -23.4 & 141.3 $\pm$ 1.3 & 247.0 $\pm$ 2.4 & 72.4 $\pm$ 9.8 &  -18.7 $\pm$ 0.1 & 2.0 $\pm$ 0.3 & 3.4 $\pm$ 0.5 \\
    L4 & 122 & -24.0 & 90.2 $\pm$ 2.0 & 244.5 $\pm$ 5.6 & 13.0 $\pm$ 0.1 & -21.2 $\pm$ 0.01 &6.9 $\pm$ 0.2 & 18.8 $\pm$ 0.5 \\
    L5 & 15 & -25.9 & 29.5 $\pm$ 2.0 & 51.3 $\pm$ 3.5 & 46.1 $\pm$ 2.7 &  -21.7 $\pm$ 0.06 & 0.6 $\pm$ 0.1 & 1.1 $\pm$ 0.1 \\
    L6 & 120 & -25.7 & 24.5 $\pm$ 2.3 & 62.9 $\pm$ 5.9 & 17.2 $\pm$ 16.0 &  -22.6 $\pm$ 1.0 & 1.4 $\pm$ 1.3 & 3.7 $\pm$ 3.4 \\
    \enddata
    \caption{Overview of our sample's UV and  \Lya\ properties. The objects' rest-frame \Lya\ equivalent width are taken from \citet{Gladders2002} and \citet{Bayliss2011,Bayliss2014}. The ultraviolet absolute magnitudes were calculated from our best-fit SED models. The \Lya\ fluxes and luminosities were calculated from the HST narrowband filter imaging following the methodology reported in \autoref{subsec:Photometry}. We report the flux in ergs/s/cm$^2$. We report the luminosities in ergs/s. The magnifications were calculated from the lens models of each system. We note that L5 has an \Lya\ EW of 15, slightly below the common \Lya\ EW threshold of 20. We include it in our sample due to its large observed \Lya\ flux and high signal-to-noise.}
    \label{tab:lyaprop}
\end{deluxetable*}

\subsection{HST Narrow-band Imaging} \label{subsec:NB}
We obtained narrow-band imaging isolating \Lya\ for each object in our sample taken with the narrow-band ramp filters installed on Hubble's Advanced Camera for Surveys (ACS). The reduced science images are available on the Harvard Dataverse \citet[10.7910/DVN/9Q0YYW]{NB_DOI}. These filters are tunable, allowing for coverage of precise wavelength ranges. We obtained these data as part of HST guest observer (GO) program \#13639 (PI: Bayliss). Each ramp filter provides an image with a narrow bandwidth ($\Delta \lambda / \lambda \simeq 2$\%) over a field of view covering approximately $\approx 40\arcsec \times 60\arcsec$ on the sky. Each source received between 1 and 3 orbits of ramp filter observations ($\sim2400-7800$s total integration time). We list the ramp filter used, the tuned central wavelength, and exposure time for each lensed LAE in Table~\ref{tab:filters}. Figure~\ref{fig:spec} shows the spectra of all six lensed LAEs centered on \Lya\ with the transmission curves of the ramp filters used to isolate \Lya\ for each source.  We reduced The ACS ramp filter data following standard procedures using Drizzlepac\footnote{\url{http://www.stsci.edu/scientific- community/software/drizzlepac.html}} \citep{Drizzlepac}. We drizzled the exposures taken in each filter using the \texttt{astrodrizzle} routine with a Gaussian kernel and a drop size of \texttt{final\_pixfrac} $= 0.8$. We combined them to a common world coordinate system (WCS) using \texttt{tweakreg} and \texttt{tweakback}, and co-added them with \texttt{astrodrizzle} onto a common reference grid with North up and a pixel scale of $0\farcs03$ px$^{-1}$. The final reduced narrowband images for each LAE are background limited.

\subsection{Broadband Imaging} \label{subsec:Broadband}

We also analyze broadband imaging of each of our lensed LAEs from several different 
observatories, spanning the optical through infra-red. For L1 and L4 we use  
i- and z-band magnitudes measured with the GMOS-N instrument on the Gemini-North 
telescope and published by \citet{Bayliss2010}. We measure additional broadband imaging 
photometry from HST and Spitzer data for all six lensed LAEs.

\subsubsection{HST}

Five of our lensed LAEs---L1 through L5--- were imaged with the Wide Field Camera 3 (WFC3) 
on the Hubble Space Telescope using both the IR and UVIS channels as a part of HST GO 
program \# 13003 (PI: Gladders). In short, each field was observed in two UVIS broadbands---one 
of [F814W, F775W, F606W], and 
one of [F475W, F390W]---and in two IR broadbands---F160W and one of [F125W, F110W, F105W]. 
L4 is a slight exception, as it has UVIS imaging only in F390W, as well as imaging from Hubble's 
Wide Field Planetary Camera 2 (WFPC2) in F606W and F814W that was taken as a part of HST
GO program \# 11974 (PI: Allam). All of these observations are described in \citet{Sharon2020}, and 
we use the same reductions described there. 

L6 has HST imaging from several different programs, all of which we incorporate into our analysis. 
The available data include: WFPC2/F814W (13200 s) and WFPC2/F606W (6600 s) obtained by GO \# 9135 (PI: Gladders), ACS/F850LP (1949 s) obtained by GO \# 13639 (PI: Bayliss), 
and ACS/F814W (8046 s), WFC3-IR/F125W (12486 s), WFC3-IR/F160W (15369 s) from GO \# 14497 (PI: Smit). We reduced these data using the same procedure that was applied to the ACS ramp 
filter data described above. For a more detailed description of the reduction methods see \citet{Sharon2020}. 

The available broadband data \citep{SGAS_DOI} allows us to fit spectral energy distributions with multiple points over a broad 
wavelength range for our entire sample, always with at least two HST broadband filters that sample the stellar 
continuum emission redward (and uncontaminated by) \Lya.

\subsubsection{Spitzer}

Our lensed LAE sample has rest-frame optical broadband imaging available from 
observations taken with the Infrared Array Camera (IRAC) on the Spitzer Space Telescope. The reduced science images are available on the Harvard Dataverse \citep[10.7910/DVN/JSVGM8]{Spitzer_DOI}.
For sources L1 - L5  we have IRAC imaging in Channel 1 (Ch1; $\sim$3.6$\mu$m) and Channel 2 
(Ch2; $\sim$4.5$\mu$m) taken as a part of programs \# 60158 (PI: Gladders), \# 70154 (PI: Gladders), 
and \# 90232 (PI: Rigby). For L6 we have IRAC imaging in Ch1, Ch2, Channel 3 (Ch3; $\sim$5.8$\mu$m), and 
Channel 4 (Ch4; $\sim$8$\mu$m), taken as a part of program \# 20754 (PI: Ellingson). Ultimately we only use 
the Ch1 and Ch2 data of L6 in our analysis because the combination of larger IRAC point spread functions (PSF) 
and higher backgrounds in Ch3 and Ch4 prevent us from meaningfully constraining the flux in those bands. 
The L6 Spitzer imaging was reduced with the same procedure as described in section 3.2 of \citet{Florian2021} with AOR 15102976. The Spitzer data sample the rest-frame $\sim$6000-9000\AA\ spectra of our lensed sources. 

To correct the small coordinate offset ($\sim1\farcs5$) between the HST and Spitzer data, we used Source Extractor \citep{SExtractor} catalogs of the HST and Spitzer imaging to match positions of bright objects in the fields, and registered the data set onto the same WCS reference frame.

\section{Methodology}
\subsection{Photometry} \label{subsec:Photometry}

AB magnitudes were measured in every available broadband filter, and are shown in \autoref{tab:phot}. The photometry was performed using elliptical apertures that match well to isophotal contours of the LAEs. We then applied PSF-dependent encircled energy (HST) and aperture correction (Spitzer) calibrations in each band to avoid aperture-based color effects. When determining the calibration values with elliptical apertures, we treated the minor axis as an effective radius. Distortion from gravitational lensing occurs along an angular path, with which we aligned the major axis of our elliptical apertures. The less distorted radial direction is thus aligned with the minor axis, and better represents the true angular size of the image.

We estimated the error in the photometry as a combination of the background noise and poisson noise. To obtain the background noise we took measurements of the blank sky nearby each LAE, masking out bright sources. The standard deviations of these pixel sets were taken as the background noise in each pixel. The total background noise was then $\sqrt{N}*\sigma_{sky}$, where N is the number of pixels within the photometric aperture. The Poisson noise was calculated as $\sqrt{N_e}$ where $N_e$ is the (background subtracted) number of electrons measured within the photometric aperture. These errors were summed in quadrature to obtain the total error reported.

We report the narrowband \Lya\ flux and luminosity in \autoref{tab:lyaprop}. The \Lya\ flux density was calculated in the same manner as the broadband photometry described above. The total flux was calculated by multiplying the flux density by the bandwidth of the narrowband filter it was observed in. We remove the continuum contribution to the narrowband flux for each object. We do this by fitting the continuum both redward and blueward of \Lya\ (see \autoref{fig:spec}) and using the continuum fit to directly compute the fraction of the total integrated flux in each narrowband filter that results from the continuum vs line emission. 
We compute the \Lya\ luminosity from the continuum-subtracted \Lya\ flux and the cosmological luminosity distance at the LAE redshift. We do not use the \Lya\ photometry in our SED modeling because \texttt{Prospector} does not rigorously model \Lya\ radiative transfer. 

Multiple imaging refers to when one lensed object has multiple known images. For objects in our sample that were multiply imaged we analyzed only a subset, preferentially choosing the brightest and/or most isolated. The chosen images are the ones indicated by the yellow bars in \autoref{fig:view}. L1, L2, L3, and L5 have multiple images, but only one was bright enough and isolated enough from other nearby bright sources. L4, while highly magnified, is not multiply imaged. L6 has three bright, isolated images. Our photometry of L6 is a combination of all its images, as they are not separable in the Spitzer imaging.

It is visually obvious from the panel containing L5 in \autoref{fig:view} that the image contains three bright regions. However, only one of these regions contains \Lya\ (the westernmost region, see \autoref{fig:roguesgallery}). Like L6, our photometry of L5 is a combination of all three regions because they are not separable in the Spitzer imaging. However, our offset measurements (\autoref{subsec:Dist}) consider only the region containing \Lya.

Many of the LAEs in our sample have significantly brighter neighboring sources which could in principle contaminate the photometry. This is especially a concern in the Spitzer data as Spitzer pixels and PSF are large compared to the LAE sizes and the HST PSF, leading to blending of sources. To correct for this, the software GALFIT \citep{Galfit} was used to subtract out the bright nearby sources. The GALFIT models were Sérsics, 2D Gaussians, or a linear combination of the two. Models of the PSF were empirically created from the available stars in each field and photometric band. The geometric models of the galaxy were convolved with the appropriate PSF model in each iteration of the GALFIT algorithm.

L6 has a bright spot initially believed to be a bright star-forming image of the LAE. However, spectral analysis by \citet{Swinbank2007} shows that it is actually a separate galaxy at a redshift of 3.66 that happens to coincide with the lensed arc. This object is labeled by the cyan bars in Figure \ref{fig:view}, and was subtracted using GALFIT before any photometric measurements were performed. In the Spitzer data where this galaxy is indistinguishable from the LAE, we attempt to subtract it by first galfitting a model to it in F160W. We then transfer the F160W model to the Spitzer data, keeping the position and morphological parameters fixed while allowing the brightness to change.

Our GALFIT models fit the outer regions of galaxies well, but left behind residuals in the cores. However, in most cases these residuals were ignored. Through iterating GALFIT, we were able to build models where the residuals were small and spatially removed enough from our apertures. In a subset of the Spitzer imaging this was not the case, and our apertures contained the imperfect residuals. To combat contamination in these cases we calculated photometry given five GALFIT models with different model parameterization and similarly good quality of fit, then report the average. Particularly in L1-L4, this led to higher uncertainties than due to the sky background alone. L1-L4 are spatially compact and the most susceptible to blending with their bright neighbors. The magnitude of the GALFIT systematic uncertainties was similar to the magnitude of the uncertainty from the sky background in any particular realization. In all cases, the GALFIT residuals are believed to be a consequence of imperfect PSF modeling and/or contamination from intra-cluster light. 

\newpage

\subsection{Stellar Population Synthesis Modelling with \texttt{Prospector}} \label{subsec:Prospector}

We utilize the \texttt{Prospector} \citep{Prospector} framework for Bayesian Spectral Energy Distribution (SED) fitting and stellar population synthesis modeling based on the photometric and spectral measurements reported in this work. \texttt{Prospector} utilizes stellar libraries \citep{FSPS,FSPS2}, and employs Monte-Carlo Markov Chains \citep[MCMC;][]{Emcee} to sample posterior distributions of galaxy parameter spaces. Each parameter in this multi-dimensional space corresponds one-to-one with galaxy properties such as stellar mass or $A_{V}$. 

\begin{figure*}[!t]
    \includegraphics[width=0.99\textwidth]{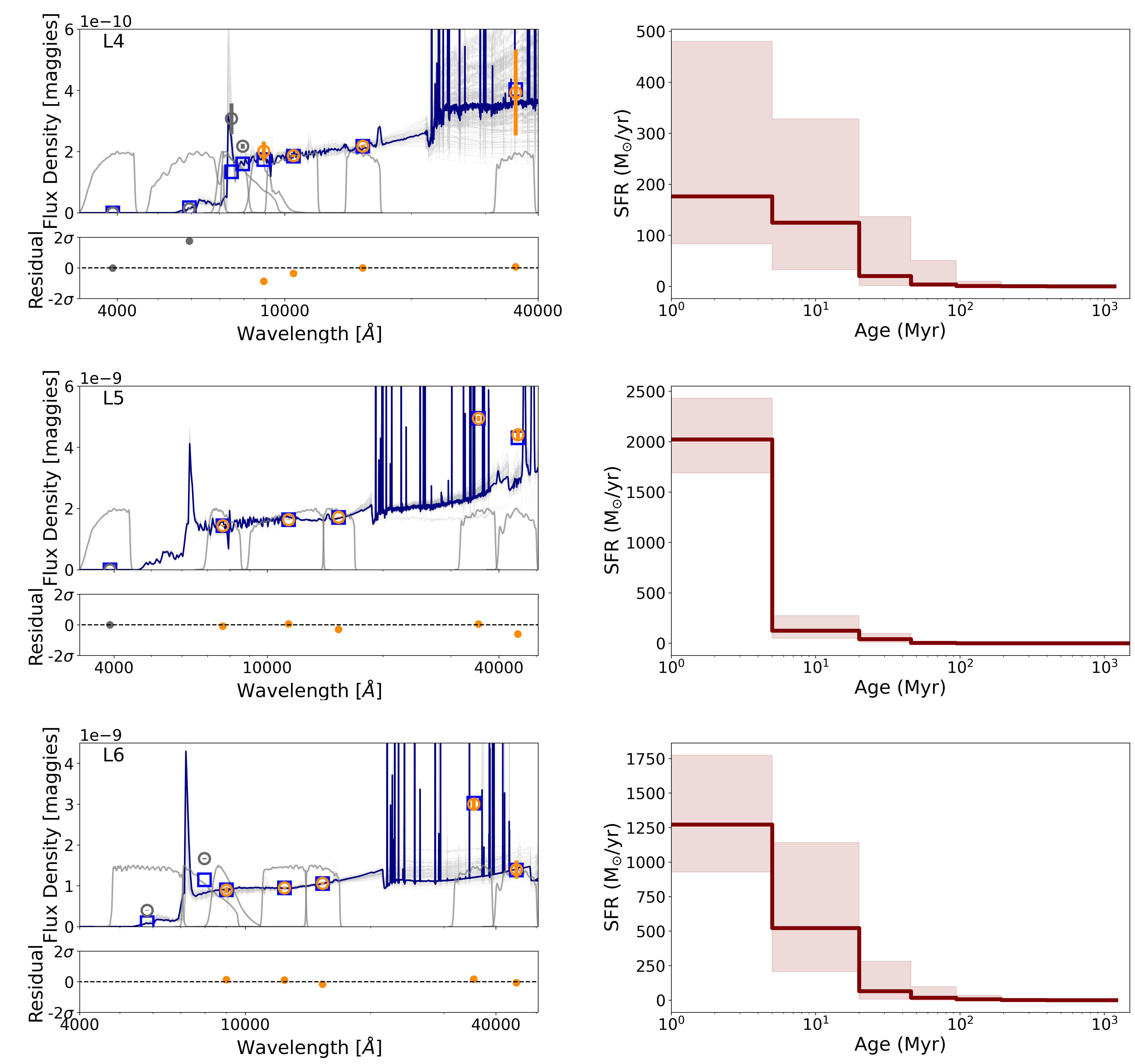}
    \figcaption{\textbf{Left:} SED models. The best fit model for each LAE is shown by the dark blue curve. The gray curves behind the best fit curve are the final 100 iterations from the end of the Emcee chain. These approximate the spread of models \texttt{Prospector} fits to the data. Filter transmission curves for our photometry are plotted at the bottom of each plot in gray. They are shifted in wavelength space to match the redshift of the LAE and are aligned with the photometric points. \textbf{Right:} Non-parametric star-formation histories. Age is defined as the lookback time from the time we currently see to the age of the universe at the object's redshift. The line and shaded regions are the 16th, 50th, and 84th percentiles of star formation rate in each age bin. The SFRs are calculated in the image plane and are hence magnified. \label{fig:SED}}
\end{figure*}

The best fit SEDs for our sample of LAEs are shown in Figure \ref{fig:SED}. 
The reported properties and uncertainties are the 16th, 50th, and 84th percentiles of these distributions. These percentiles correspond to the mean and standard deviations of a Gaussian distribution. While the probability distributions sampled by \texttt{Prospector} are not necessarily Gaussian, these percentiles still provide a suitable measure of the values and uncertainties.

We ran \texttt{Prospector} with a combination of the parametric\_sfh, dust\_emission, and nebular libraries. These libraries allow us to jointly model stellar mass, stellar metallicity, stellar population age, amount of dust, gas-phase metallicity, and ionization parameter U as free parameters. However, the constraining power of our photometry only robustly constrained the stellar mass. The amount of dust and age of the stellar population are less constrained, but are reported for context. The stellar and gas phase metallicites and the ionization parameter U were not meaningfully constrained. Redshift was kept fixed in all cases except for L4, which had a strong H$\alpha$ emission line on the edge of the IRAC Channel 1 transmission curve. This was inferred from the brightness of the \Lya\ line and the \Lya-based redshift. We do not have a precise measurement of the true systematic redshift, so in this case we allowed the redshift to vary over a small range calculated using \citet{Verhamme2018}: an investigation into offsets between systematic and \Lya\ based redshift measurements. The extent to which H$\alpha$ lies under the transmission curve has a large impact on the inferred properties from the fit. Allowing redshift to vary for L4 ensures we are not biased towards only one possible set of parameters. Additionally, we chose to exclude any photometry from bands that contained \Lya\ because \texttt{Prospector} cannot model \Lya\ emission rigorously.

\begin{figure*}[!b]
    \centering
    \includegraphics[width=0.99\textwidth]{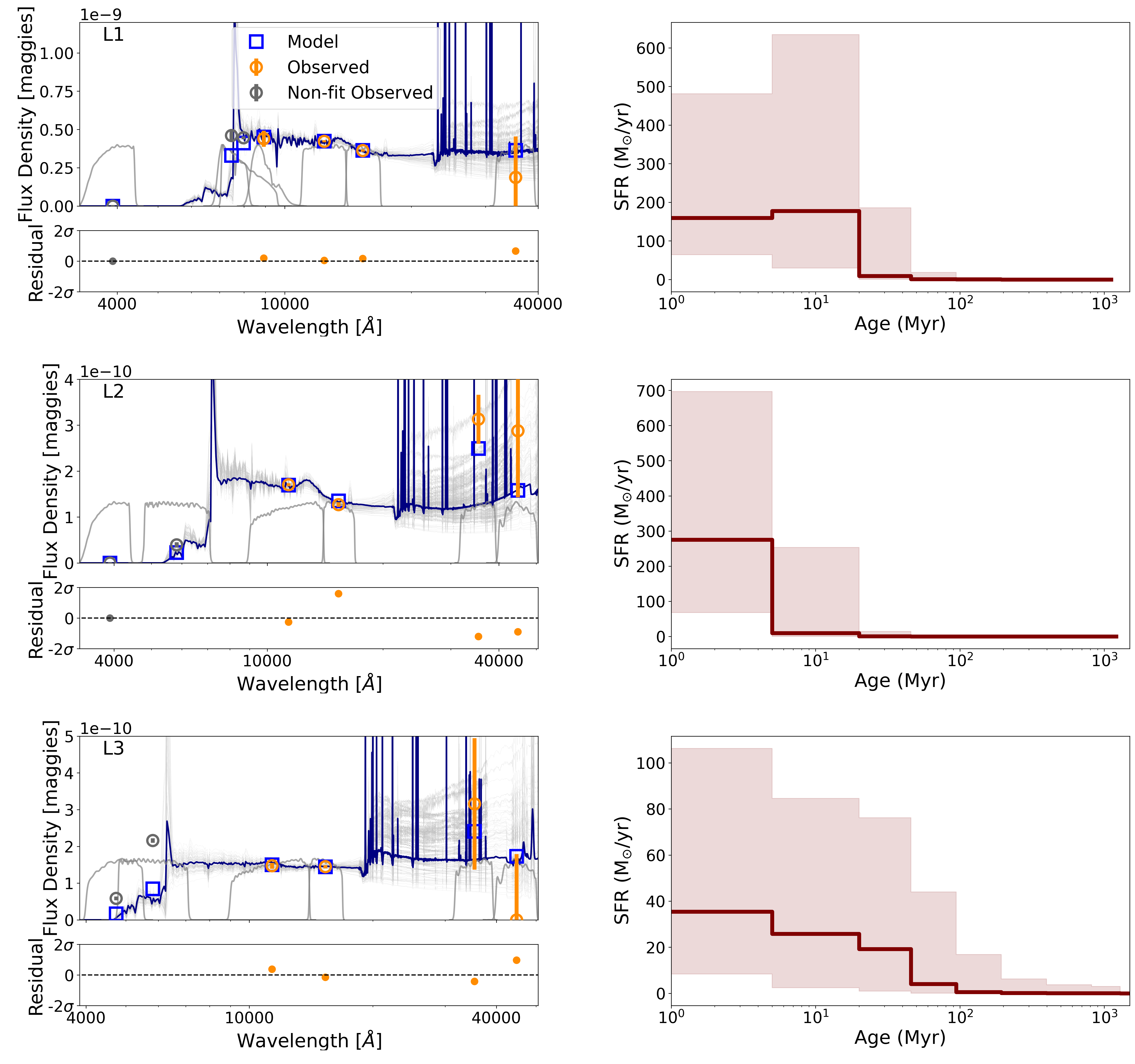}
\end{figure*}

At the redshifts of our sample the HST photometry constrains the UV slope and luminosity of the young stellar populations. The Spitzer/IRAC data sample broad swaths of the rest-frame optical for all of our sources and measures the continuum emission from older stellar populations and nebular emission lines in each lensed LAE. The Spitzer data therefore provide constraints on the strength of the Balmer jump, marginalized over uncertainties in the strengths of the nebular emission lines.

We make use of \texttt{Prospector}'s ability to model non-parametric SFHs; we chose a non-parametric SFH with fixed time bins and the continuity prior. As described in \citet{Leja2019}, this SFH setup is the best case scenario for recovering true SFHs with photometry-only SED fitting. We define eight age bins:
\begin{center}
    0 < t $\le$ 10 Myr\\
    10 < t $\le$ 30 Myr\\
    30 < t $\le$ 60 Myr\\
    60 < t $\le$ 130 Myr\\
    130 < t $\le$ 260 Myr\\
    260 < t $\le$ 550 Myr\\
    550 Myr < t $\le$ 1.1 Gyr\\
    1.1 Gyr < t $\le$ t$_{final}$\\
\end{center}
where t$_{final}$ is the age of the universe at the redshift of the LAE. Except for the first two and last age bins, the age bins are separated equally in logarithmic time, following the methodology of \citet{Leja2019}. The first two bins were chosen to probe populations of very young (<10 Myr and 10-30 Myr old) stars. Each LAE was modeled with 256 walkers and a total of 9216 MCMC iterations, excluding three preliminary "burn-in" runs.

\begin{deluxetable*}{c c c c c c c c c}[ht]
    \startdata
    \\
    ID & \Lya\ &log(Stellar Mass) & log(Stellar Mass) & Mass-Weighted & $t_{50}$ & $t_{90}$ & Dust1  & Dust2\\
     & Morphology & [$M_{\odot}$] (Lensed) & [$M_{\odot}$] (Intrinsic) & Age [Myr] & [Myr] & [Myr] & [$A_{V}$] & [$A_{V}$] \\
    \hline\hline
    L1 & Clumpy &$9.85^{+0.62}_{-0.48}$ & $8.15^{+0.64}_{-0.51}$ & $17^{+19}_{-8}$ & $13^{+14}_{-7}$ & $44^{+510}_{-25}$ & $0.10^{+0.23}_{-0.08}$ & $0.82^{+0.92}_{-0.69}$\\
    L2 & Clumpy &$9.72^{+0.43}_{-0.61}$ & $8.60^{+0.44}_{-0.61}$ & $8^{+9}_{-3}$ & $4^{+11}_{-1}$ & $18^{+220}_{-13}$ & $0.09^{+0.14}_{-0.07}$ & $2.15^{+0.64}_{-1.30}$\\
    L3 & Extended &$9.69^{+0.58}_{-0.68}$ & $7.79^{+0.58}_{-0.68}$ & $33^{+83}_{-20}$ & $38^{+230}_{-26}$ & $187^{+850}_{-150}$ & $0.57^{+0.84}_{-0.43}$ & $0.35^{+0.75}_{-0.29}$\\
    L4 & Extended & $9.94^{+0.45}_{-0.40}$ & $8.82^{+0.45}_{-0.40}$ & $20^{+27}_{-9}$ & $17^{+41}_{-7}$ & $87^{+580}_{-68}$ & $0.46^{+0.47}_{-0.28}$ & $0.54^{+0.69}_{-0.39}$\\
    L5 & Clumpy &$10.41^{+0.09}_{-0.08}$ & $8.71^{+0.10}_{-0.10}$ & $7^{+2}_{-1}$ & $4^{+1}_{-1}$ & $33^{+12}_{-18}$ &$0.23^{+0.06}_{-0.06}$ & $1.00^{+0.16}_{-0.15}$\\
    L6 & Extended & $10.55^{+0.26}_{-0.26}$ & $9.24^{+0.47}_{-0.37}$ & $16^{+11}_{-06}$ & $14^{+26}_{-6}$ & $130^{+490}_{-110}$ & $0.42^{+0.56}_{-0.31}$ & $0.80^{+0.39}_{-0.44}$\\
    \enddata
    \caption{Table of inferred galaxy properties from \texttt{Prospector}. The associated SEDs are found in \autoref{fig:SED} and correspond with the above quantities. Dust is broken into two components: dust1 and dust2. These are defined as the amount of dust attenuation affecting stars with ages $<10$ Myr (dust1) and $>10$ Myr (dust2). This distinction is based on the work of \citet{Charlot2000}, where the environments of young and old stellar populations contain different amounts of dust (see \autoref{subsec:Prospect Analysis}). We report three measures of the age: mass-weighted age, $t_{50}$, and $t_{90}$, as defined in \autoref{subsec:Prospector}. We include the \Lya\ morphology classification (see  \autoref{subsec:Morphology}) for reference.}
    \label{tab:prop}
\end{deluxetable*}

We report three measures of the age of the stellar population: mass-weighted age, $t_{50}$, and $t_{90}$. The mass-weighted age is defined as the integral of the SFH weighted by the total amount of mass formed in each time bin. $t_{50}$ and $t_{90}$ are defined as the times at which the (unweighted) integral of the SFH is equal to 50\% and 90\% of the total stellar mass formed. 

The errors on all three quantities are drawn from 1000 realizations of the SFH (1000 different walker/iteration combinations).From these distributions we report the 16th, 50th, and 84th percentiles. It is critical to note that \texttt{Prospector} calculates the stellar mass posterior distribution from image plane photometry, which is magnified by strong gravitational lensing. We calculate the intrinsic stellar mass distribution by convolving the posterior with a Gaussian distribution of the object's magnification factor.

\subsection{Lens Models}
Modeling the gravitational potential of the lensing system is an important aspect of analyzing strongly lensed sources. Creation of magnification maps, identification of all the images of a lensed source, and creation of source-plane images through ray tracing are all examples of obtaining information only available through lens modeling. In this work we used magnification maps to calculate the intrinsic stellar mass and intrinsic \Lya\ magnitudes of our sample. We also identified previously unknown images of L1 and L5, but they were too faint for robust analysis. We leave use of lens models to calculate physical (source plane) sizes and distances for the subsequent paper on this sample Navarre et al. (in prep).  

Lensing models of the systems were created using \texttt{Lenstool} \citep{Jullo_2007}, which is a parametric lens-modeling software in which projected mass density halos are linearly combined. All models assumed a pseudo-isothermal ellipsoidal mass distribution (PIEMD, \citet{Limousin2007}) with the following parameters: position, ellipticity, position angle, core radius, truncation radius, and normalization. The models are iteratively created; beginning with modeling the most obvious evidence of lensing, then using the output to identify additional constraints. 

Lens models for L1 through L5 were computed by \citet{Sharon2020}. The lens model for L6 was created for this work. All of the lens model outputs are publicly available through MAST\footnote{\url{https://archive.stsci.edu/hlsp/sgas}}. Here we provide only a short summary, and refer the reader to \citet{Sharon2020} for more details. The details of the lens model of L6 will be presented in (Navarre et al. (in prep)).

\newpage

\subsection{Offset Measurements} \label{subsec:Dist}
Through visual analysis of the stellar continuum and narrowband \Lya\ imaging, we found separate bright regions within the images of L1, L2, and L5. We denote these regions with lowercase letters (L1a, L2b, etc.). Note these should not be misinterpreted as multiple images of the same object, like those found L6. The positions of each region are shown in \autoref{tab:position} and can be seen visually by the Xs in \autoref{fig:roguesgallery}.

We calculated the on-sky offset between the stellar continuum and \Lya\ emission of each region of interest. To represent the stellar continuum, we chose the bluest filter available that contained the UV slope and did not contain \Lya. We calculated the on-sky positions by using DS9's \citep{DS9} centroiding algorithm in each band.

\subsection{Classification of Clumpy vs. Extended} \label{subsec:CvsE}
\begin{figure*}[!htb]
    \centering
    \includegraphics[scale=0.6]{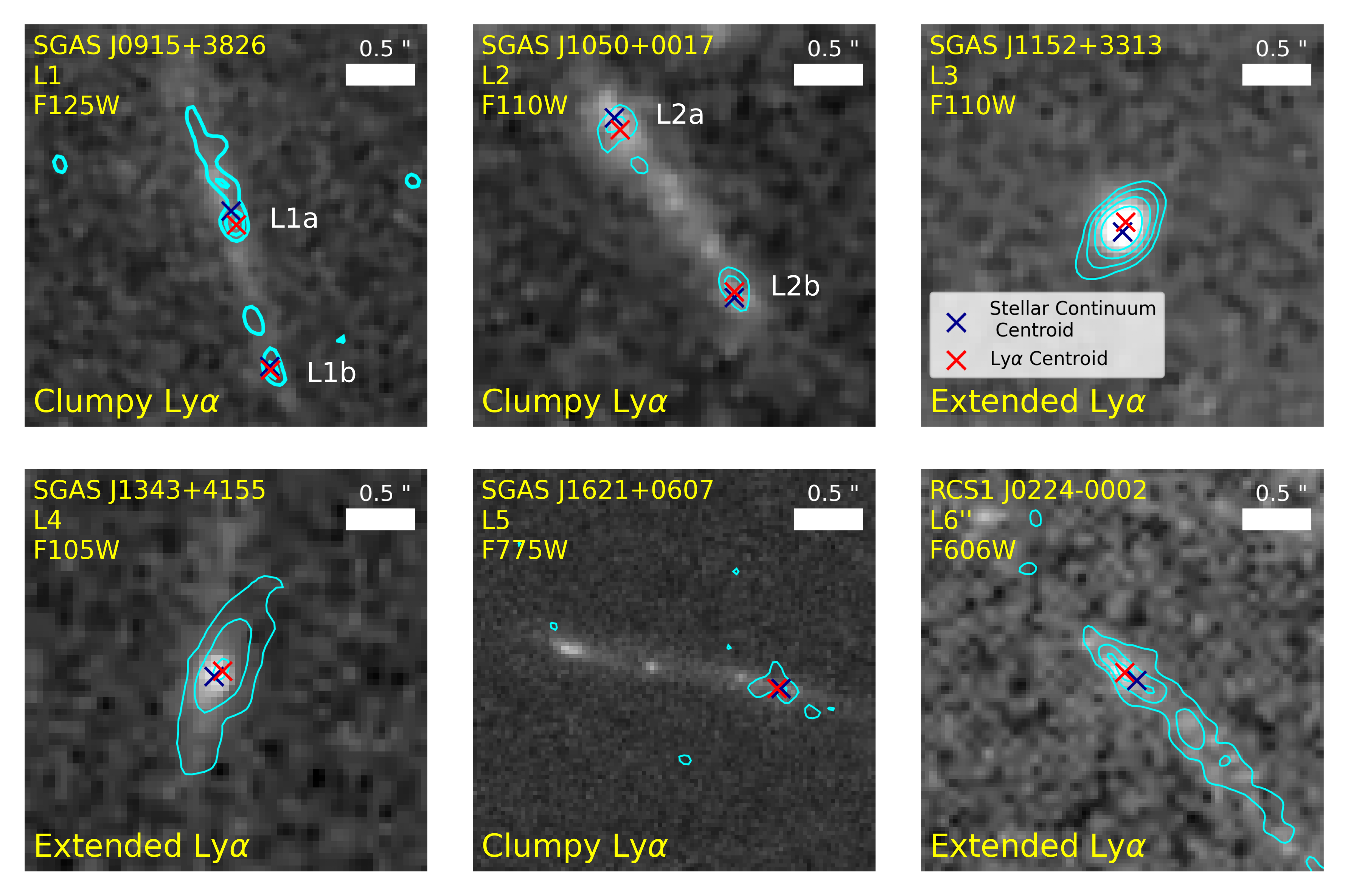}
    \figcaption{View of the LAE morphologies in the stellar continuum (grayscale) and Lyman-alpha (cyan contours). The centroids of each clump are marked with blue and red Xs , corresponding with stellar continuum and \Lya\ respectively. We explicitly label the different bright regions found within L1 and L2 in white lettering. The sigma levels of the \Lya\ contours are as follows: L1:[2,3,4], L2:[2,3], L3:[2,4,8,16], L4:[2,4,8], L5:[1,2], L6:[2,4,6]. \label{fig:roguesgallery}} 
\end{figure*}

We classify the \Lya\ morphology of our LAEs in two broad categories: clumpy and extended. This classification is derived from the spatially-resolved ratio of the \Lya\ emission to the UV continuum emission. Our observations in both \Lya\ and the UV continuum are background limited, and we interpret their pixel-by-pixel ratio as something akin to a \Lya\ escape fraction map. The UV continuum is created from the same young, hot stars that create \Lya. Additionally, the ionizing photon emission from these stars should scale with the UV continuum. Assuming a $f_{esc,LyC} \sim 0$, every ionizing photon should result in approximately 0.68 \Lya\ photons (from \citet{Dijkstra2017}). In some LAEs we see both regions of high \Lya, low UV; and regions of low \Lya, high UV surface brightness. This can be seen visually in \autoref{fig:roguesgallery}. We classify these LAEs as clumpy. We classify the other LAEs as extended, as their \Lya\ surface brightness scales more smoothly with the UV continuum into the CGM.

\begin{figure}[!htb]
    \centering
    \includegraphics[scale=0.58]{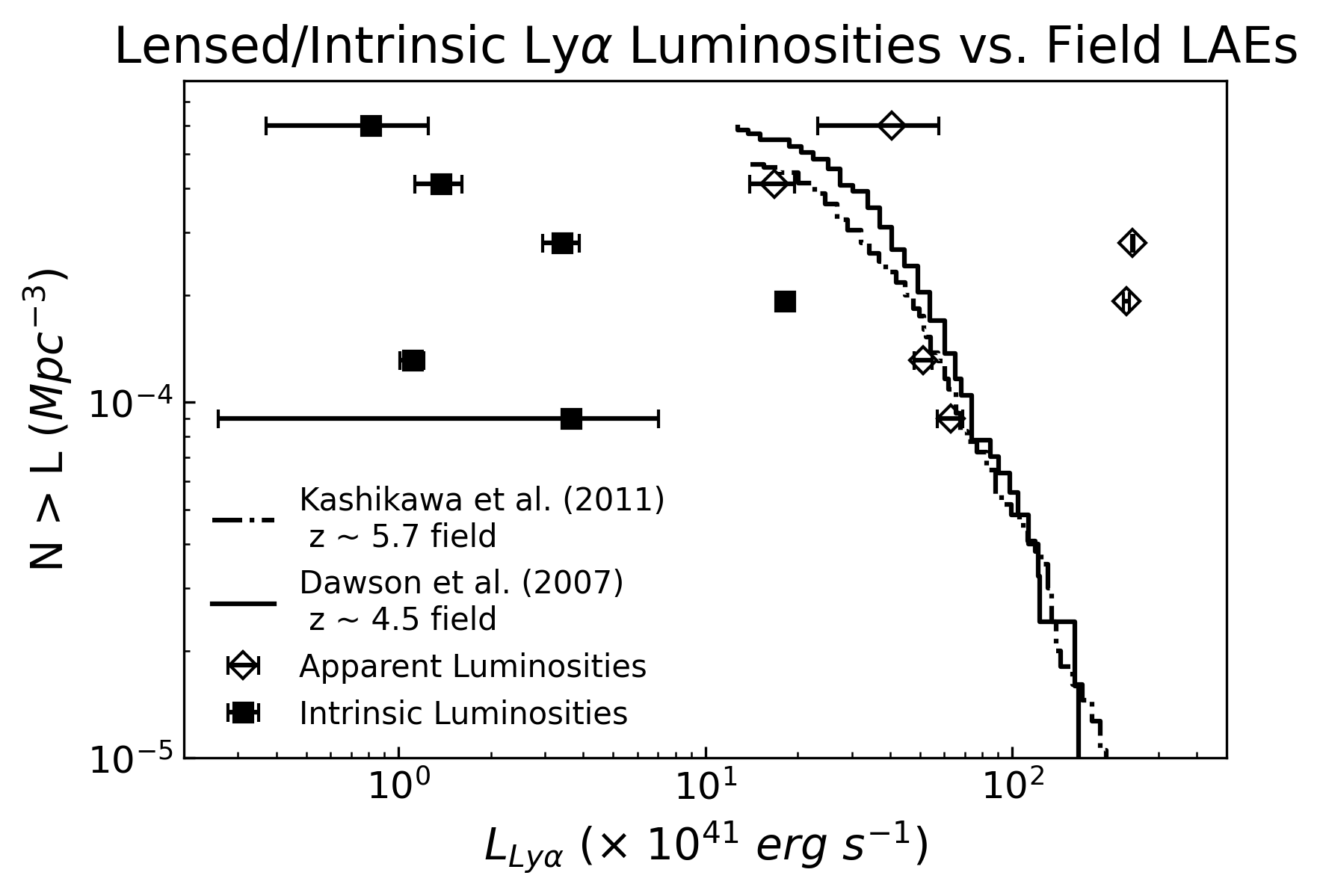}
    \figcaption{The solid and dashed lines show luminosity densities of high-redshift field LAEs from \citet{Kashikawa2011} and \citet{Dawson2007}. The open diamonds are the measured \Lya\ luminosities of our sample. The filled squares are the intrinsic (delensed) luminosities, calculated from our lens model magnification maps. Note that the y-coordinates of our data points do not represent a number density. The intrinsic and observed luminosities of an LAE are placed at the same y-coordinate for clarity. Our observed luminosities are consistent with or brighter than typical field LAEs at these redshifts. The y-axis ordering of the points corresponds with L1 at the top and descends to L6 at the bottom.\label{fig:brightness}}
    
\end{figure}

\section{Analysis} \label{sec:Analysis}

\subsection{Inferred Galaxy Properties} \label{subsec:Prospect Analysis}
Due to the strong lensing magnification our lensed LAE galaxies have apparent brightnesses that are among the brightest LAEs discovered in blank field narrow-band surveys similar redshifts, but are intrinsically much fainter than typical field LAEs (\autoref{fig:brightness}) . 
In \autoref{tab:prop} we report the inferred stellar mass, age of the stellar population, and dust content ($A_{V}$). We compare our results with the following studies: \citet{Chary2005,Gawiser2006,Lai2007,Pirzkal2007,Finkelstein2009,Hayes2013,Santos2021}. 

Our intrinsic stellar mass values are consistent with the other studies which find high-redshift LAEs having stellar masses $\sim 1\times10^7 M_{\odot} - 8.5\times10^9 M_{\odot}$. Our sample probes the low mass end of the $z \sim 4-5$ LAE mass functions \citep{Santos2021}. This is unsurprising given that gravitational lensing is a rare phenomenon and low mass LAEs are more numerous than high mass LAEs. 

Our age measurements correspond with young stellar populations: $t_{50} \lesssim 30$ Myr. Young stellar populations create large amounts of \Lya, which is consistent with their high \Lya\ equivalent widths. The studies we compare against report ages $\sim 5 - 850$ Myr inferred through SED fitting. However, these values are obtained by fitting simple stellar population (SSP) or exponentially-decaying (tau-model) SFHs. \citet{Carnall2019} shows that use of these star-formation models imposes strong priors on physical parameters such as stellar population age. Our age measurements were inferred using non-parametric SFHs, which return galaxy properties with less bias and more accuracy \citep{Leja2019}. We find that the clumpy LAEs tend to be younger while the extended LAEs tend to be older, which we discuss further in \autoref{subsec:Morphology}.

We break dust attenuation into two categories: attenuation affecting young stellar populations ($<10$ Myr, dust1) and attenuation affecting all older stellar populations ($>10$ Myr, dust2). This distinction is based on the work of \citet{Charlot2000}. Physically, the attenuation in the birth clouds of young, massive stars can be different than the attenuation in regions where older stars have migrated away from their birth clouds. Older stellar populations may have had their surrounding gas disrupted through feedback from younger, more energetic stars. The studies to which we compare our data report dust extinctions from emission line diagnostics or SED fitting with only one dust component. 

Our estimations for attenuation from young stellar populations (dust1) are consistent with these studies: having $A_V \sim 0-1$. We find that the LAEs with extended \Lya\ morphologies have larger $A_V$ than those with clumpy \Lya. Since dust destroys \Lya\ photons, it is counter-intuitive to find more in systems with extended \Lya. Our results imply that the main contributor to a clumpy vs. extended \Lya\ morphology is the H\,I distribution rather than the amount of dust.

\begin{deluxetable}{c c c}[ht]
    \startdata
    \\
    ID & Image Plane $\delta_{Ly\alpha}$ [$\arcsec$] & Stellar Continuum Band \\
    \hline\hline
    L1a & 0.090 & F125W \\
    L1b & 0.042 & F125W \\
    L2a & 0.067 & F110W \\
    L2b & 0.030 & F110W \\
    L3 & 0.060 & F110W \\
    L4 & 0.042 & F105W \\
    L5 & 0.030 & F775W \\
    L6$'$ & 0.095 & F850LP \\
    L6$''$ & 0.108 & F850LP \\
    L6$'''$ & 0.095 & F850LP \\
    \enddata
    \caption{Image-plane (on-sky) angular offsets between the stellar continuum and \Lya\ emission. We list the filters used for the stellar continuum, which were chosen as the bluest available filters redward of \Lya\ that did not contain \Lya. Since our measurements do not correct for distortion from gravitational lensing, our findings should be considered as upper limits to the intrinsic (source plane) separation.}
    \label{tab:offsets}
\end{deluxetable}

\subsection{Ly-\texorpdfstring{$\alpha$}{a} and UV Spatial Coincidence} \label{subsec:Spatial Coincidence}
The measured image plane spatial separation between the \Lya\ centroid and stellar continuum is reported in \autoref{tab:offsets} as $\delta_{Lya}$. Studies measuring large samples of field LAEs indicate that a significant portion contain a spatial offset. \citet{Jiang2013} found that there is little offset between UV and \Lya\ positions in compact galaxies ($<$ 0\farcs2), but merging/interacting systems can be significantly offset ($>$ 0\farcs3). \citet{Shibuya2014} found the majority of their LAE sample is offset by $<$ 0\farcs2, yet only $\approx 23\%$ are classified as mergers. Our measurements do not take into account the distortion from gravitational lensing and should be considered upper limits to the true (source plane) separations. We find that all of the \Lya\ regions studied in our sample have small offsets $<$ 0\farcs2, indicating that they are likely not merging nor interacting galaxies.

\subsection{Ly-\texorpdfstring{$\alpha$}{a} Morphology} \label{subsec:Morphology}
Robust quantitative descriptions of the \Lya\ morphology require forward modeling through the lensing potential, and is left to the followup paper Navarre et al. (in prep.). Here we present a qualitative discussion of the morphologies. One can see in \autoref{fig:roguesgallery} that the \Lya\ contours fall into two broad categories: extended and clumpy. L1, L2, and L5 have clumpy \Lya\ morphologies: the \Lya\ emission does not cover the entire galaxy, instead being concentrated in specific regions. L3, L4, and L6 have extended morphologies: their \Lya\ emission covers the entire galaxy and extends beyond the associated stellar continuum emission. This broadening is consistent with \Lya\ radiative transfer.

We propose that our LAEs containing clumpy \Lya\ morphologies have varying H\,I column densities across their star-forming regions. Since H\,I scatters \Lya, lines of sight with lower H\,I column densities preferentially allow more \Lya\ photons to escape. The \Lya\ clumps seen in our sample could align with these lines of sight. This hypothesis is supported by previous observations of \Lya\ emitting galaxies with evidence of non-uniform H\,I distributions \citep{Heckman2011,RV2017,Chisholm2018,Gazagnes2018,Steidel2018}. Furthermore, simulations of a similar galaxy in \citet{Blaizot2023} find significant anisotropic effects on the \Lya\ line shape and luminosity. These effects correlate with gas flows and evolve over short time scales ($\sim$ tens Myr), implying the possibility of short-lived gas configurations that can create clumpy \Lya\ morphologies.

Our sample suggests a progression from clumpy \Lya\ to extended \Lya\ over a time scale of $\sim 14-38$ Myr based on $t_{50}$. The clumpy LAEs tend to be younger while the extended LAEs tend to be older. We note that L1 and L6 have similar ages while being classified differently, and explain this through our SFH resolution. Our first three SFH age bins cover (0,10] Myr, (10,30] Myr, and (30,61.7] Myr respectively. The upper error in $t_{50}$ of L6 is larger than that of L1, extending beyond 30 Myr while L1 does not. This implies that the stellar population of L6 contains more older stars than that of L1. Due to the errors on the ages and our small sample size we do not claim that this observation is robust; instead merely suggestive.

A possible explanation of this apparent difference is that clumpy \Lya\ morphologies appear in a short-lived stage of starbursts. In this initial stage young massive stars ionize channels along preferential lines of sight. As more stars are formed and greater densities of ionizing photons are created, more lines of sight are punched through the ISM. These ionized channels allow \Lya\ to more easily directly escape the LAE through them into the IGM, explaining the presence of the clumpy \Lya\ morphologies. It is also possible that the inflowing gas that fuels bursts of star formation obscures lines of sight that would otherwise leak \Lya\ \citep{Blaizot2023}. However, not all \Lya\ that escapes a star forming region necessarily immediately exits into the IGM. Ionized channels allow \Lya\ to escape into the extended halo of neutral hydrogen that surrounds the LAE, where it can then scatter into an extended profile.     

\section{Summary} \label{sec:Conclusion}
We have investigated the image-plane properties of six \Lya\ emitting galaxies at $4.1 < z < 5.2$. Using multi-band imaging from HST and Spitzer, we were able to infer stellar mass, ages, and $A_V$s with \texttt{Prospector}. Taking advantage of magnification from gravitational lensing and HST resolution narrowband filters we were able to measure \Lya\ on sub-galactic scales otherwise inaccessible. With the HST narrowband \Lya\ imaging we were able to see different types of \Lya\ morphology: clumpy vs. extended. Through use of lens models we were able to calculate the intrinsic stellar mass and \Lya\ luminosity. Our findings are summarized below.

\begin{itemize}
  \item The intrinsic stellar masses, ages, and amounts of dust are consistent with values from other studies of high and low redshift LAEs. In particular, the young ages of the stellar populations are consistent with the kinds of stars that produce high amounts of \Lya.
  \item The offsets between the stellar continuum and \Lya\ emission of our sample are small ($<0\farcs2$). This is consistent with other studies which find small spatial offsets, and can be explained through radiative transfer through non-uniform H\,I distributions. The small offsets indicate our LAEs are not interacting nor merging systems.
  \item A qualitative analysis of the \Lya\ emission shows two broad categories of \Lya\ morphologies: clumpy and extended. We find that the LAEs containing clumpy \Lya\ generally have younger stellar populations. This suggests a possible progression from clumpy \Lya\ to extended \Lya, perhaps driven by the growth of young massive stars and creation of ionized channels during a starburst period. 
\end{itemize}

Future work (Navarre et al. in prep) will further study the source-plane (intrinsic) morphologies of this sample. That work will include state-of-the-art custom forward modeling code to robustly measure sizes and distances through the lensing potential. It will measure physical sizes of the \Lya\ regions, stellar continuum regions, and \Lya\ offsets; reconstruct the source-plane images of the LAEs, and detail the new lens model for L6. We will directly compare the physical information with samples of low and high redshift LAEs, as well as with the inferred galaxy properties from this paper. We will characterize the variations in the morphology of the \Lya\ emission and compare them to low-redshift LAEs, where the distribution of \Lya\ is heterogeneous. \citep{Bond2010,Finkelstein2011,Steidel2011,Feldmeier2013,Momose2014}

\section*{Acknowledgements}
Support for HST program HST-GO-13639 was provided by NASA through a grant from the Space Telescope Science Institute, which is operated by the Association of Universities for Research in Astronomy, Inc., under NASA contract NAS 5--26555.

Based on observations made with the NASA/ESA Hubble Space Telescope, obtained at the Space Telescope Science Institute, which is operated by the Association of Universities for Research in Astronomy, Inc., under NASA contract NAS 5-26555. These observations are associated with programs HST-GO-9135, HST-GO-11974, HST-GO-13003, HST-GO-13639, HST-GO-14497. 

This work is based in part on observations made with the Spitzer Space Telescope, which was operated by the Jet Propulsion Laboratory, California Institute of Technology under a contract with NASA. Support for the Spitzer programs used in this work was provided by NASA. Spitzer data used in this analysis were taken as a part of programs \# 20754, \# 60158, \# 70154, and \# 90232.

Some of the data presented in this paper were obtained from the Multimission Archive at the Space Telescope Science Institute (MAST): \dataset[10.17909/jvg9-v026]{http://dx.doi.org/10.17909/jvg9-v026}. 

\clearpage

\bibliography{refs.bib}{}
\bibliographystyle{aasjournal}

\end{document}